\documentclass[aps,prl,twocolumn]{revtex4-1}
\usepackage{graphicx}
\usepackage{graphics}
\usepackage{braket} 
\usepackage{epstopdf}
\usepackage{dcolumn}
\usepackage{latexsym} 
\usepackage{amssymb}
\usepackage{amsmath}
\usepackage{amsfonts}
\usepackage{color}
\usepackage{bbold}

\begin{document}


\title{Anisotropic Optical Shock Waves in Isotropic Media with Giant Nonlocal Nonlinearity}

\author{Giulia Marcucci}
\email{giulia.marcucci@uniroma1.it}
\affiliation{Department of Physics, University Sapienza, Piazzale Aldo Moro 2, 00185 Rome (Italy)}
\affiliation{Institute for Complex Systems, Via dei Taurini 19, 00185 Rome (Italy)}
\author{Phillip Cala}
\affiliation{Department of Physics and Astronomy, San Francisco State University, San Francisco, California 94132, USA}
\author{Weining Man}
\affiliation{Department of Physics and Astronomy, San Francisco State University, San Francisco, California 94132, USA}
\author{Davide Pierangeli}
\affiliation{Department of Physics, University Sapienza, Piazzale Aldo Moro 2, 00185 Rome (Italy)}
\affiliation{Institute for Complex Systems, Via dei Taurini 19, 00185 Rome (Italy) }
\author{Claudio Conti}
\email{claudio.conti@uniroma1.it}
\affiliation{Institute for Complex Systems, Via dei Taurini 19, 00185 Rome (Italy)}
\affiliation{Department of Physics, University Sapienza, Piazzale Aldo Moro 2, 00185 Rome (Italy)}
\affiliation{TEDA Applied Physics Institute, Nankai University, Tianjin 300457 (P. R. China)}
\author{Zhigang Chen}
\email{zhigang@sfsu.edu}
\affiliation{Department of Physics and Astronomy, San Francisco State University, San Francisco, California 94132, USA}
\affiliation{TEDA Applied Physics Institute, Nankai University, Tianjin 300457 (P. R. China)}

\date{\today}

\begin{abstract}

Dispersive shock waves in thermal optical media belong to the third-order nonlinear phenomena, whose intrinsic irreversibility is described by time asymmetric quantum mechanics.
Recent studies demonstrated that nonlocal wave breaking evolves in an exponentially decaying dynamics ruled by the reversed harmonic oscillator, namely, the simplest irreversible quantum system in the rigged Hilbert spaces. The generalization of this theory to more complex scenarios is still an open question.
In this work, we use a thermal third-order medium with an unprecedented giant Kerr coefficient, the M-Cresol/Nylon mixed solution, to access an extremely-nonlinear highly-nonlocal regime and realize anisotropic shock waves. We prove that a superposition of the Gamow vectors in an ad hoc rigged Hilbert space describes the nonlinear beam propagation beyond the shock point. Specifically, the resulting rigged Hilbert space is a tensorial product between the reversed and the standard harmonic oscillators spaces.
The anisotropy turns out from the interaction of trapping and antitrapping potentials in perpendicular directions.
Our work opens the way to a complete description of novel intriguing shock phenomena, and those mediated by extreme nonlinearities.

\end{abstract}


\maketitle


Dispersive shock waves~(DSWs) are widespread nonlinear phenomena in physics, from hydrodynamics~\cite{1966Peregrine,1988Smyth,2016Maiden} to acoustics~\cite{1970Taylor}, from Bose-Einstein condensates~\cite{2004Damski,2004Kamchatnov,2004Perez,2005Simula,2006Hoefer,2008Chang} to plasma physics~\cite{2008Romagnani}, and from quantum liquids~\cite{2006Bettelheim} to optics~\cite{1967Akhmanov,1989Rothenberg,2007El,2007HoeferPhysD,2007Ghofraniha,2007Wan,2008Conti,2012Ghofraniha,2013Garnier,2013Gentilini,2014Gentilini,2014Smith1,2015GentiliniPRA,2015GentiliniSciRep,2015Xu,2016Braidotti,2016Wetzel,2016XuMussot,2016XuGarnier,2017Zannotti,2019Gautam,2019Marcucci,2019MarcucciReview}.

In optics and photonics, when light propagates in a medium whose refractive index depending on the beam intensity, (e.g., light experiences the Kerr effect~\cite{2008Boyd}), the interplay between diffraction (or dispersion) and nonlinearity can lead to steep gradients in the phase profile, and, in some cases, to a wave breaking~\cite{2019MarcucciReview}.
Such discontinuity is regularized by very rapid oscillations, both in phase chirp and in intensity outlines, called undular bores~\cite{2007Wan}.
If we slightly mutate this third-order nonlinearity, making it nonlocal (or noninstantaneous), then the phenomenology changes very little. When the phase chirp reaches the discontinuity and starts to oscillate, the intensity does not develop undular bores, but rather the annular collapse singularity~(ACS) (or M-shaped singularity if the beam is $1+1$-dimensional)~\cite{2015Xu,2016Braidotti,2018Xu,2019MarcucciReview}. Spatial collapse-DSWs occur in thermal media, where the radiation-matter interaction is led by a thermo-optic effect, and the refractive index perturbation depends on the whole intensity profile~\cite{2008Boyd,2019MarcucciReview}.

Theoretically, such modification to the Kerr nonlinearity has significant consequences. Laser beam propagation in a standard Kerr medium is ruled by the nonlinear Schr\"odinger equation~(NLSE), exactly solvable by the inverse scattering transform method~\cite{1967Gardner,1972Zacharov,2015Fibich}. However, the NLSE with a nonlocal potential cannot be solved by inverse scattering transform (despite some recent progress in two-dimensional~(2D) media~\cite{2017Hokiris,2019Hokiris}), but only through the Whitham modulation and the hydrodynamic approximation~\cite{1999Whitham}. Moreover, the dynamics beyond the shock appears to be intrinsically irreversible, as recently shown by applying the time asymmetric quantum mechanics~\cite{1981Bohm,1989Bohm,1998Bohm,1999Bohm,2002Chruscinski,2002Delamadrid,2003Gadella,2003Chruscinski,2004Chruscinski,2004Civitarese,2016Celeghini,2016Marcucci} to the description of DSWs in highly nonlocal approximation~\cite{2015GentiliniPRA,2015GentiliniSciRep,2016Braidotti,2017Marcucci,2019MarcucciReview}.

In highly nonlocal Kerr media, the ACS is modeled by the simplest Hamiltonian of time asymmetric quantum mechanics: the reversed harmonic oscillator~(RHO). This Hamiltonian has a real complete continuous spectrum, which corresponds to a basis of eigenfunctions that have simple poles in their analytical prolongations to the complex plane. Since the RHO is a harmonic oscillator~(HO) with a pure imaginary frequency, starting from the HO complete point spectrum, it turns out that the RHO point spectrum is the set of the mentioned simple poles, and the related eigenfunctions are non-normalizable eigenvectors belonging to a rigged Hilbert space, called Gamow vectors~(GVs)~\cite{1928GamowDE,1928GamowENG}.
Light propagation beyond the collapse is then expressed as a superposition of GVs, which exponentially decay with quantized decay rates. Such laser beam evolution is the outcome of a phenomenon, the shock, that is intrinsically irreversible: in the absence of absorption and interaction with an external thermal bath, the dynamics cannot be inverted, i.e., it is time asymmetric.
Can this theoretical model be used to describe much more complex scenarios? To answer this question, we need to access regimes with much stronger nonlinearity.

In recent experiments, it is showed that M-Cresol/Nylon solutions exhibit an isotropic giant self-defocusing nonlinearity, tunable by varying the nylon concentration~\cite{2014Smith}.
M-Cresol/Nylon is a thermal chemical mixture, consisting of an organic solvent (m-cresol) and a synthetic polymeric solute (nylon).
The nonlinear Kerr coefficient $n_2$ of pure m-cresol is $-9\times10^{-8} cm^2/W$, but it was found that, in such mixtures, $n_2=-1.6\times10^{-5} cm^2/W$ for a nylon mass concentration of $3.5\%$, higher than other thermal nonlinear materials in which ACSs have been observed~\cite{2014Smith1,2019MarcucciReview}.

In this Letter, we report on our theoretical discovery and experimental evidence of optical DSWs with an \textit{anisotropic zero-singularity}~(ZS) (i.e., a gap in the intensity profile along only one direction) in M-Cresol/Nylon solutions. Fixing $z$ as the longitudinal and $x,y$ as the transverse directions, we consider an initial beam which is even in the $y$ direction, and odd along the $x$ direction. This initial condition causes a new phenomenon: the shock does develop an annular collapse, but around the ZS it presents an abrupt intensity discontinuity.
We theoretically analyze this anisotropic wave breaking. We model the beam propagation beyond the shock point by time asymmetric quantum mechanics and uncover the mechanism that determines how such an abrupt intensity discontinuity is generated. We numerically simulate these results and find remarkable agreement between experimental outcomes and theoretical predictions.


For a laser beam propagating in a thermal medium with refractive index $n=n_0+\Delta n[|A|^2](\mathbf{R})$, where $\mathbf{R}=(\mathbf{R_{\perp}},Z)=(X,Y,Z)$, the NLSE describes the evolution of the envelope $A(\mathbf{R})$ of the monochromatic field $\mathbf{E}(\mathbf{R})=\hat{\mathbf{E}}_0 A(\mathbf{R})e^{\imath k Z}$, and it reads as follows:
\begin{equation}
2\imath k \partial_Z A+\nabla_{\mathbf{R_{\perp}}}^2 A+2k^2 \frac{\Delta n[|A|^2]}{n_0} A=-\imath\frac{k}{ L_{loss}}A,
\label{eq:NLSE}
\end{equation}
where $\nabla_{\mathbf{R_{\perp}}}^2=\partial_X^2+\partial_Y^2$, $k=\frac{2\pi n_0}{\lambda}$ is the wavenumber, $\lambda$ is the wavelength, and $L_{loss}$ is the linear loss length. By defining $I=|A|^2$ the intensity, $\bar{P}(Z)=\int\int\mathrm{d}\mathbf{R_{\perp}}I(\mathbf{R})$ the power, $L_d=k W_0^2$ the diffraction length, with $W_0$ the initial beam waist, and $\alpha=\frac{L_d}{L_{loss}}$, it turns out that $\bar{P}$ is not conserved only if $\alpha\neq0$. Indeed, if $\alpha\sim0$, then $\partial_Z \bar{P}\sim0$~\cite{1991Lisak}.

In low absorption regime, the refractive index perturbation in Eq.~(\ref{eq:NLSE}) is~\cite{2019MarcucciReview}
\begin{equation}
\Delta n[|A|^2](\mathbf{R_{\perp}})=n_2 \int\int\mathrm{d}\mathbf{R_{\perp}'}K(\mathbf{R_{\perp}}-\mathbf{R_{\perp}'})I(\mathbf{R_{\perp}'}),
\label{eq:n1}
\end{equation}
with $n_2$ the Kerr coefficient and $K(\mathbf{R_{\perp}})$ is the kernel function describing the nonlocal nonlinearity, normalized such that $\int\int\mathrm{d}\mathbf{R_{\perp}}K(\mathbf{R_{\perp}})=1$.
For $K(\mathbf{R_{\perp}})=\delta(\mathbf{R_{\perp}})$ we attain the well-known local Kerr effect, i.e., $n=n_0+n_2I$~\cite{2008Boyd}. In our nonlocal case, where the laser beam produces a thermo-optic effect that generates an isotropic variation of the medium density distribution, the response function is
\begin{equation}
K(X,Y)=\tilde{K}(X)\tilde{K}(Y),\;\; \tilde{K}(X)=\frac{e^{-\frac{|X|}{L_{nloc}}}}{2L_{nloc}}.
\label{eq:kernel}
\end{equation}
Here $L_{nloc}$ is the nonlocality length~\cite{1997Snyder,2003Conti,2007Minovich,2007Ghofraniha,2019Marcucci}.
We rescale Eq.~(\ref{eq:NLSE}), with $\alpha\sim0$, by defining the dimensionless variables $x=X/W_0$, $y=Y/W_0$ and $z=Z/L_d$, and obtain
\begin{equation}
  \imath \partial_z \psi+\frac{1}{2}\nabla_{\mathbf{r_{\perp}}}^2\psi+\chi P K_0*|\psi|^2 \psi=0,
\label{eq:NLSEnorm}
\end{equation}
where $\mathbf{r}=(\mathbf{r_{\perp}},z)=(x,y,z)$, $\nabla_{\mathbf{r_{\perp}}}^2=\partial_x^2+\partial_y^2$, $\psi(\mathbf{r})=\frac{W_0}{\sqrt{\bar{P}}}A(\mathbf{R})$, $\chi=\frac{n_2}{|n_2|}$ and $P=\frac{\bar{P}}{P_{REF}}$ with $P_{REF}=\frac{\lambda^2}{4\pi^2 n_0 |n_2|}$.
The asterisk in Eq. (\ref{eq:NLSEnorm}) stands for the convolution product, while $K_0(x,y)=\tilde{K_0}(x)\tilde{K_0}(y)$ with $\tilde{K_0}(x)=W_0 \tilde{K}(X)=\frac{e^{-\frac{|x|}{\sigma}}}{2\sigma}$, $\tilde{K_0}(y)$ of the same form, and $\sigma=\frac{L_{nloc}}{W_0}$ the nonlocality degree.

In highly nonlocal approximation ($\sigma>>1$), once the initial conditions are fixed, $|\psi|^2$ mimics a delta function (or a narrow superposition of delta functions), and the nonlocal potential looses its $I$-dependence, becoming a simple function of the transverse coordinates~\cite{1997Snyder,2012Folli}:
\begin{equation}
\begin{array}{l}
K_0*|\psi|^2\simeq \kappa(\mathbf{r_{\perp}})\simeq\kappa(\mathbf{0})+\left(\partial_x\kappa|_{\mathbf{r_{\perp}}=\mathbf{0}}\right)x+\left(\partial_y\kappa|_{\mathbf{r_{\perp}}=\mathbf{0}}\right)y+\\ \\
+\frac{1}{2}\left(\partial_x^2\kappa|_{\mathbf{r_{\perp}}=\mathbf{0}}\right)x^2+\left(\partial_x\partial_y\kappa|_{\mathbf{r_{\perp}}=\mathbf{0}}\right)xy+\frac{1}{2}\left(\partial_y^2\kappa|_{\mathbf{r_{\perp}}=\mathbf{0}}\right)y^2,
\end{array}
\label{eq:nonlocpot}
\end{equation}
after a Taylor second-order expansion.
This approximation maps the NLSE in Eq.~(\ref{eq:NLSEnorm}) into a linear Schr\"odinger equation $\imath \partial_z\psi(\mathbf{r})=\hat{H}(\mathbf{p_{\perp}},\mathbf{r_{\perp}})\psi(\mathbf{r})$, with $\hat{H}(\mathbf{p_{\perp}},\mathbf{r_{\perp}})=\frac{1}{2}\hat{\mathbf{p_{\perp}}}^2+\hat V(\mathbf{r_{\perp}})$ the Hamiltonian, $\mathbf{\hat p_{\perp}}=(\hat p_x, \hat p_y)=(-\imath\partial_x, -\imath\partial_y)$ the transverse momentum and $\hat V(\mathbf{r_{\perp}})=-\chi P\kappa(\mathbf{r_{\perp}})\mathbb{1}$ the multiplicative potential ($\mathbb{1}$ is the identity operator). Let us consider the initial condition
\begin{equation}
\psi_{\mathrm{ISO}}(\mathbf{r_{\perp}})=\psi_{even}(x)\psi_{even}(y),\;\;\psi_{even}(x)=\frac 1{\sqrt[4]{\pi}}e^{-\frac{x^2}2},
\label{eq:isoini}
\end{equation}
and $\psi_{even}(y)$ of the same form.
The shape of $\kappa(\mathbf{r_{\perp}})$ depends on $\psi_{\mathrm{ISO}}(\mathbf{r_{\perp}})$. Indeed, since $\psi_{\mathrm{ISO}}$ is an even, separable function, all the first derivatives in Eq.~(\ref{eq:isoini}) vanish, hence $\kappa(\mathbf{r_{\perp}})=\kappa_0^2-\frac{1}{2}\kappa_2^2\left|\mathbf{r_{\perp}}\right|^2$, where $\kappa_0^2=\frac1{4\sigma^2}$ and $\kappa_2^2=\frac{1}{2\sqrt{\pi}\sigma^3}$.
\begin{figure}[h!]
\begin{center}
\includegraphics[width=\linewidth]{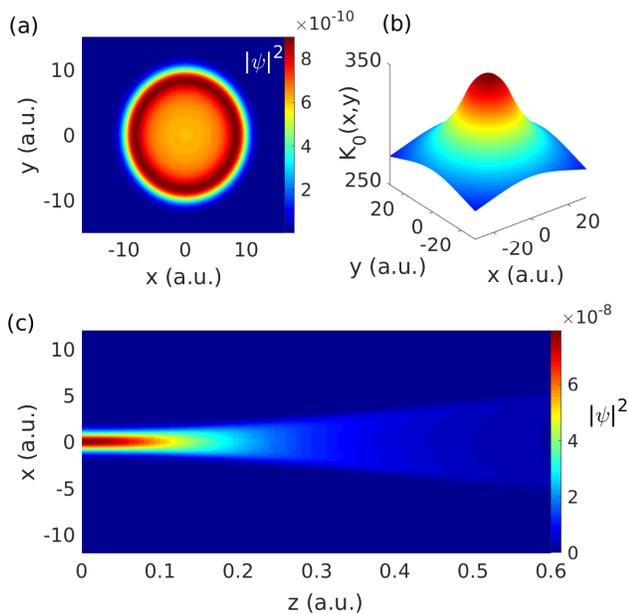}
\end{center}
\caption{Solution of the NLSE~(\ref{eq:NLSEnorm}) with initial condition~(\ref{eq:isoini}), for $P=4\times10^6$ and $\sigma=120$, in arbitrary units: (a) shows the intensity transverse profile at $z=0$, (b) exhibits the symmetric response function derived from Eq.~(\ref{eq:isoini}), and (c) reports the intensity longitudinal outline, here on the plane $(x,z)$, equal to one on the plane $(y,z)$.
\label{fig:iso}}
\end{figure}

In the defocusing case ($n_2<0$), the transverse profile of the solution of Eq.~(\ref{eq:NLSEnorm}) with initial condition~(\ref{eq:isoini}) is shown in Fig.~\ref{fig:iso}(a). Figure~\ref{fig:iso}(b) exhibits the central part of the symmetric response function $K_0(x,y)$, while the longitudinal profile on $x,z$ (same of $y,z$) is reported in Fig.~\ref{fig:iso}(c). The corresponding Hamiltonian reads
$\hat{H}=P\kappa_0^2 +\hat{H}_{\mathrm{RHO}}(p_x,x)+\hat{H}_{\mathrm{RHO}}(p_y,y)$,
where
\begin{equation}
\hat{H}_{\mathrm{RHO}}(p_x,x)=\frac 12\hat{p_x}^2-\frac{\gamma^2}{2}\hat{x}^2
\label{eq:RHO}
\end{equation}
is the 1D-RHO Hamiltonian of frequency $\gamma=\sqrt{P}\kappa_2$.
Once moved to $\phi(\mathbf{r})=e^{\imath P\kappa_0^2}\psi(\mathbf{r})$, the Schr\"odinger equation becomes
$\imath \partial_z\phi(\mathbf{r})=\left[\hat{H}_{\mathrm{RHO}}(p_x,x)+\hat{H}_{\mathrm{RHO}}(p_y,y)\right]\phi(\mathbf{r})$, which is completely separable.
In bra-ket notation
\begin{equation}
\begin{array}{l}
\imath \frac{\mathrm{d}}{\mathrm{d}z}|\phi(z)\rangle=\hat{H}_{\mathrm{ISO}}(\mathbf{p_{\perp}},\mathbf{r_{\perp}})|\phi(z)\rangle,\\ \\
\hat{H}_{\mathrm{ISO}}(\mathbf{p_{\perp}},\mathbf{r_{\perp}})=\hat{H}_{\mathrm{RHO}}(p_x,x)\otimes\mathbb{1}_y+\mathbb{1}_x\otimes\hat{H}_{\mathrm{RHO}}(p_y,y),\\ \\
|\phi(z)\rangle=|\phi_{even}(z)\rangle_x\otimes|\phi_{even}(z)\rangle_y,
\end{array}
\label{eq:iso}
\end{equation}
with $\otimes$ the tensorial product, no more explicitly written hereafter. The solution of Eq.~(\ref{eq:iso}) lives in a tensorial product between two 1D rigged Hilbert spaces. Indeed, if we consider the evolution operator $\hat{U}(z)=e^{-\imath\hat{H}z}$ such that $|\phi(z)\rangle=\hat{U}(z)|\phi(0)\rangle$, for Eq.~(\ref{eq:iso}) $|\phi(z)\rangle=e^{-\imath\hat{H}_{\mathrm{RHO}}z}|\psi_{even}\rangle_xe^{-\imath\hat{H}_{\mathrm{RHO}}z}|\psi_{even}\rangle_y$. The representation of $|\phi_{even}(z)\rangle_{x,y}=e^{-\imath\hat{H}_{\mathrm{RHO}}z}|\psi_{even}\rangle_{x,y}$ in terms of GVs was already studied and was also already demonstrated to describe 1D DSWs in thermal media~\cite{2015GentiliniPRA,2015GentiliniSciRep,2016Braidotti,2017Marcucci,2019MarcucciReview}. It is $|\phi_{even}(z)\rangle_{x,y}=|\phi_N^G(z)\rangle+|\phi_N^{BG}(z)\rangle$, with
\begin{equation}
|\phi_N^G(z)\rangle=\sum_{n=0}^N e^{-\frac{\gamma}2(2n+1)z}|\mathfrak{f}_n^-\rangle\langle \mathfrak{f}_n^+|\psi_{even}\rangle
\label{eq:GV}
\end{equation}
the decaying superposition of Gamow states $|\mathfrak{f}_n^-\rangle$, corresponding to the energy levels $E_n^{RHO}=\imath\frac{\gamma}2(2n+1)$, and $|\phi_N^{BG}(z)\rangle$ the background function, both belonging to the same 1D rigged Hilbert space.

\begin{figure}[h!]
\begin{center}
\textbf{(a)}
\raisebox{-1\height}{\includegraphics[width=0.95\linewidth]{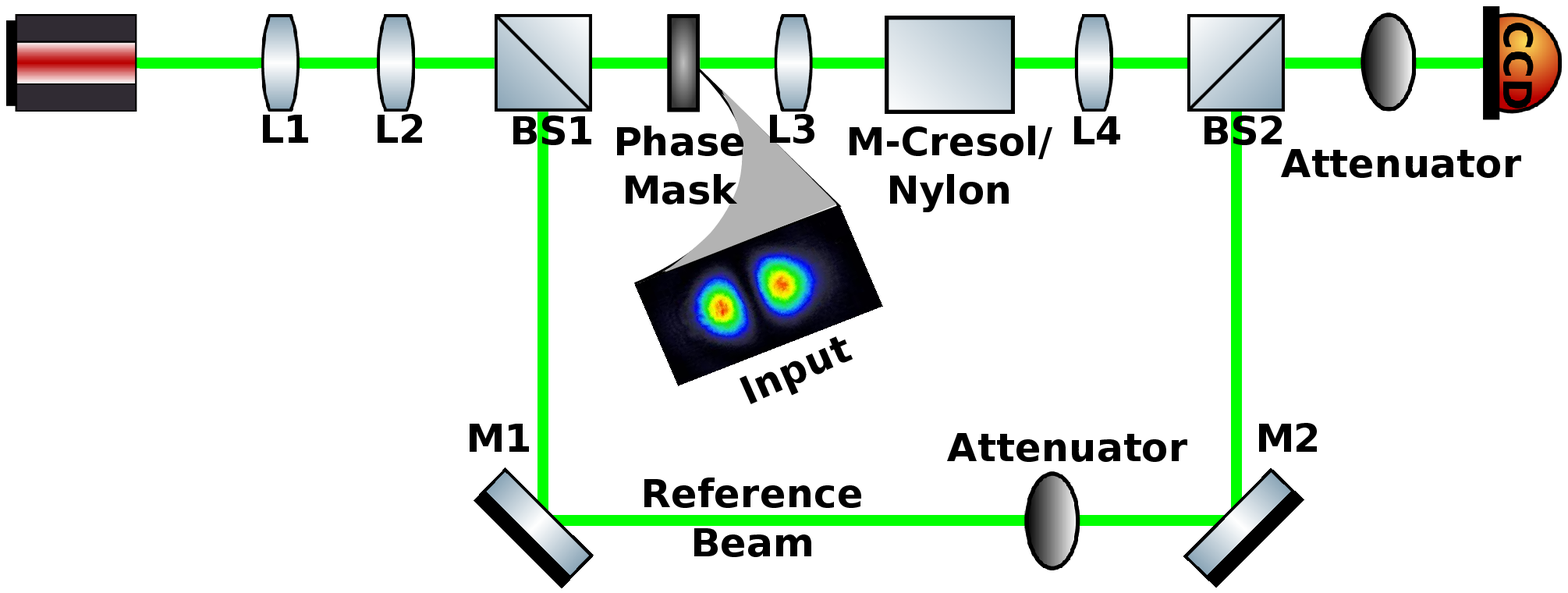}}
\\\vspace{3mm}
\textbf{(b)}\raisebox{-0.85\height}{\includegraphics[width=\linewidth]{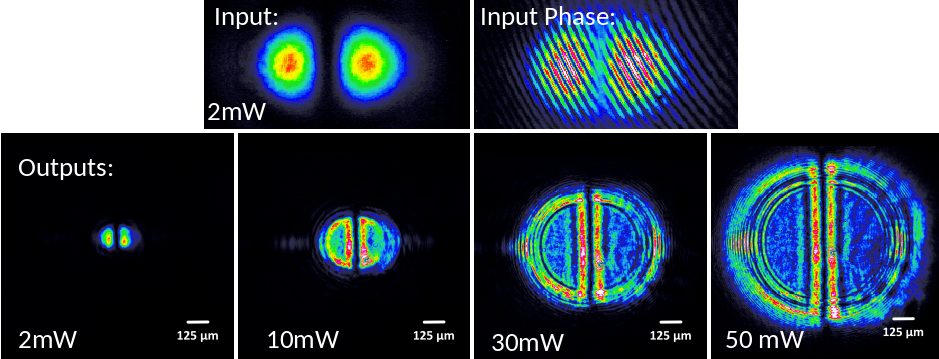}}
\end{center}
\caption{
(a) Experimental setup. A $\lambda=532$nm CW-laser beam is collimated through two lenses (L1 and L2). A beam splitter (BS1) divides the beam in two arms. the first is made asymmetric by a phase mask and propagates in a $2$mm-long cuvette filled with M-Cresol/Nylon $3.5\%$-solution. The second is a reference beam for interference measurements. The output is imaged (via L4 and BS2) onto a CCD-camera.
(b) Input and outputs observed. The phase mask in (a) generates a $\pi$ discontinuity in the input phase, here reported with initial power $\bar{P}=2$mW and waist $W_0=15.8\mu$m, together with the intensity profile. Several output at different initial powers are shown. The low power cannot distinguish nonlinear effects from diffraction, but the higher the power is, the stronger the nonlinear effects are, leading to the formation of anisotropic DSWs.
\label{fig:exp}}
\end{figure}

Our experiments are performed in an isotropic medium with asymmetric initial conditions for the beam, designed through a phase mask, to attain an anisotropic light propagation. The setup is illustrated in Fig.~\ref{fig:exp}(a). A laser beam with wavelength $\lambda=532$nm passes through two collimating lenses (L1 and L2), and then through a beam splitter (BS1), which divides the beam into two arms, one used for the nonlinear experiment, and the other for getting a reference beam for interference measurements. The beam outcoming from the first arm is transformed into an asymmetric input by a phase mask, and then is focused (via L3) onto the facet of a $2$mm-long cuvette, which contains a M-Cresol/Nylon solution  (with $3.5\%$ Nylon). The output is imaged (via L4 and BS2) onto a CCD-camera.

Figure~\ref{fig:exp}(b) reports the input beam (zoom-in intensity and phase patterns at initial power $\bar{P}=2$mW and waist $W_0=15.8\mu$m) and its outputs at different initial powers. The input beam presents a phase discontinuity of $\pi$ along $x=0$.
The output beam exhibits diffraction at a low power, but evolves into an overall shock pattern with two parts at high powers. It's important to note that, while the whole pattern expands, the gap between two parts remains constant. This represents the first realization of what we define as \textit{anisotropic DSWs}: ACSs with an initial ZS, which generates two barriers of light intensity around a constant gap in the middle of the beam. Despite the medium isotropy, the oddity of the initial condition generates an anisotropic final transverse profile.

We model the initial asymmetric beam-shape as follows:
\begin{equation}
\psi_{\mathrm{ANI}}(\mathbf{r_{\perp}})=\psi_{odd}(x)\psi_{even}(y),\;\;\psi_{odd}(x)=-\frac{\sqrt{2}}{\sqrt[4]{\pi}}xe^{-\frac{x^2}2},
\label{eq:aniini}
\end{equation}
$\psi_{even}(y)=\frac 1{\sqrt[4]{\pi}}e^{-\frac{y^2}2}$ as in Eq.~(\ref{eq:isoini}).
In this case, Eq.~(\ref{eq:nonlocpot}) is reduced to $\kappa(\mathbf{r_{\perp}})=\kappa_0^2+\frac{1}{2}\kappa_1^2 x^2-\frac{1}{2}\kappa_2^2 y^2$, with $\kappa_0^2=\frac1{4\sigma^2}$, $\kappa_1^2=\frac{1}{4\sigma^4}$ and $\kappa_2^2=\frac{1}{2\sqrt{\pi}\sigma^3}$.
\begin{figure}[h!]
\begin{center}
\includegraphics[width=\linewidth]{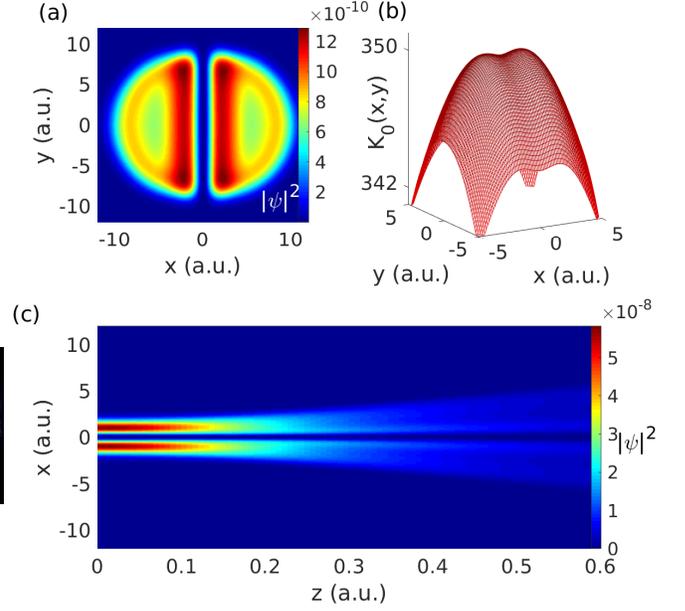}
\end{center}
\caption{
Solution of the NLSE~(\ref{eq:NLSEnorm}) with initial condition~(\ref{eq:aniini}), for $P=4\times10^6$ and $\sigma=120$, in arbitrary units: (a) shows the intensity transverse profile at $z=1$, (b) exhibits the asymmetric response function derived from Eq.~(\ref{eq:isoini}), and (c) reports the intensity longitudinal outline on the plane $(x,z)$, with the zero-singularity.
\label{fig:ani}}
\end{figure}

The anisotropy appears evident: not only the initial condition presents a ZS, but also the response function has two different behaviors along $x,y$ directions. Numerical simulations are illustrated in Fig.~\ref{fig:ani}. Figure~\ref{fig:ani}(a) shows the anisotropic DSWs, solution of the NLSE~(\ref{eq:NLSEnorm}) with initial condition~(\ref{eq:aniini}). Figure~\ref{fig:ani}(b) gives numerical proof of the response function anisotropy: the $(x,y)$-plane origin corresponds to a saddle point, with a locally increasing profile along $x>0$, $y<0$ and a locally decreasing outline along $x<0$, $y>0$. Figure~\ref{fig:ani}(c) reports the intensity-ZS in a neighborhood of $x=0$ during propagation.

The presence of the saddle point in the response function has direct consequences through highly nonlocal approximation in mapping the NLSE in the quantum-like linear Schr\"odinger equation. From the expression of $\kappa(\mathbf{r_{\perp}})$ above, for $\phi(\mathbf{r})=e^{\imath P\kappa_0^2}\psi(\mathbf{r})$ we obtain
\begin{equation}
\begin{array}{l}
\imath \frac{\mathrm{d}}{\mathrm{d}z}|\phi(z)\rangle=\hat{H}_{\mathrm{ANI}}(\mathbf{p_{\perp}},\mathbf{r_{\perp}})|\phi(z)\rangle,\\ \\
\hat{H}_{\mathrm{ANI}}(\mathbf{p_{\perp}},\mathbf{r_{\perp}})=\hat{H}_{\mathrm{HO}}(p_x,x)\mathbb{1}_y+\mathbb{1}_x\hat{H}_{\mathrm{RHO}}(p_y,y),\\ \\
|\phi(z)\rangle=|\phi_{odd}(z)\rangle_x|\phi_{even}(z)\rangle_y,
\end{array}
\label{eq:ani}
\end{equation}
where $\hat{H}_{\mathrm{HO}}(p_x,x)=\frac 12\hat{p_x}^2+\frac{\omega^2}{2}\hat{x}^2$ is the one-dimensional harmonic oscillator Hamiltonian with $\omega=\sqrt{P}\kappa_{odd}$, and $\hat{H}_{\mathrm{RHO}}$ is the one-dimensional RHO Hamiltonian in Eq.~(\ref{eq:RHO}).
The solution of Eq.~(\ref{eq:ani}) is the tensorial product of $|\phi_{odd}(z)\rangle_{x}=\sum_{n=0}^{+\infty} e^{\imath\frac{\omega}2(2n+1)z}|\Psi_n^{HO}\rangle\langle\Psi_n^{HO}|\psi_{odd}\rangle$, where $|\Psi_n^{HO}\rangle$ are $\hat{H}_{\mathrm{HO}}$-eigenstates corresponding to the energy levels $E_n^{HO}=\frac{\omega}2(2n+1)$~\cite{2016Marcucci}, and $|\phi_{even}(z)\rangle_{y}=|\phi_N^G(z)\rangle+|\phi_N^{BG}(z)\rangle$, explicitly written in Eq.~(\ref{eq:GV}).
\begin{figure}[h!]
\begin{center}
\includegraphics[width=\linewidth]{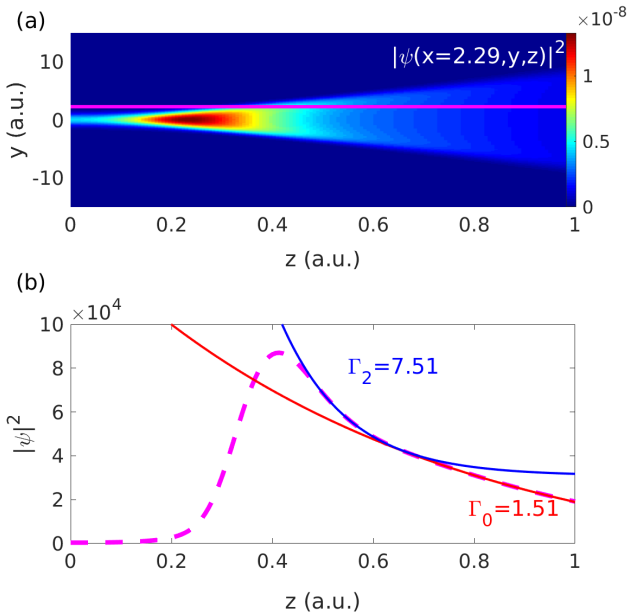}
\end{center}
\caption{
GVs signature. From Fig.~\ref{fig:ani}, in the same conditions, (a) is the $y,z$ profile at fixed $x=2.29$. Intensity along the pink line, i.e., $|\phi(x=2.29,y=2.29,z)|^2=\langle\phi_{even}(z)|\phi_{even}(z)\rangle_{y}=\sum_{n=0}^Ne^{-\Gamma_n z}\left|\langle\mathfrak{f}_n^+|\phi_{even}\rangle\right|^2$, is in (b) [$\Gamma_n=\gamma(2n+1)$ quantized decay rates].  The continuous lines represent the two first exponential functions of the summation above that fit the decaying part: the red line is the fundamental GV, with decay rate $\Gamma_0$, and the blue line is the first excited GV, with decay rate $\Gamma_2$.
\label{fig:GV}}
\end{figure}

Evidence of the presence of GVs is given in Fig.~\ref{fig:GV}. By defining $\Gamma_n=\gamma(2n+1)$, we look for the first two quantized decay rates $\Gamma_{0,2}$ (the even Gaussian initial function only leads to even energy levels) in the longitudinal propagation in $y$-direction. Indeed, if one computes the intensity of the $y$-part, one finds $\langle\phi_{even}(z)|\phi_{even}(z)\rangle_{y}\stackrel{N>>0}{\simeq}\langle\phi_N^G(z)|\phi_N^G(z)\rangle=\sum_{n=0}^Ne^{-\Gamma_n z}\left|\langle\mathfrak{f}_n^+|\phi_{even}\rangle\right|^2$. Figure~\ref{fig:GV}(a) shows the theoretical section of the nonlinear sample where we seek decaying states. We fix $x=2.29$, a little distant from the shock-gap, and report the corresponding intensity in $y,z$ plane. The pink line is equivalent to $x=y=2.29$. Figure~\ref{fig:GV}(b) exhibits $|\phi(x=2.29,y=2.29,z)|^2$, exponentially decaying. Two exponential fits demonstrate the GV occurrence: the fundamental Gamow state represents the plateau with decay rate $\Gamma_0=1.51$, whereas the first excited one interpolates the peak, with decay rate $\Gamma_2=7.51$.
We stress that the rule $\frac{\Gamma_2}{\Gamma_0}=5$ is respected.

\begin{figure}[h!]
\begin{center}
\includegraphics[width=\linewidth]{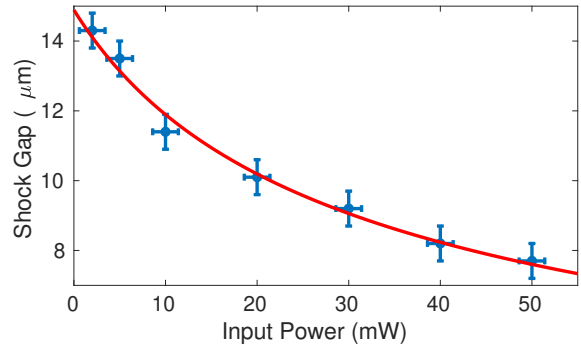}
\end{center}
\caption{
Observation of the shock-gap versus the initial power $\bar{P}$. The red line shows the theoretical fit with a functions $\propto\frac{1}{\sqrt{\bar{P}}}$.
\label{fig:gap}}
\end{figure}

The analysis of the barriers, due to the HO component, and the corresponding shock-gap is also examinated. Since $\hat{H}_{\mathrm{HO}}$ has potential $\hat{V}_{\mathrm{HO}}(x)=\frac{\omega^2}2\hat{x}^2$, we expect a shock-gap with the same behavior of the potential width $\Delta x\propto\frac1{\sqrt{\omega}}=\frac{2\sigma^2}{\sqrt{P}}$, that is, a shock-gap that varies among different experiments by changing the initial power but, in a single observation with a given $P$, does not change by varying $z$ [as shown in Fig.~\ref{fig:ani}(c)].
This is experimentally proven and reported in Fig.~\ref{fig:gap}, which shows measurements of the shock-gap at variance of initial power $\bar{P}$. A theoretical fit with a function $\propto\frac{1}{\sqrt{\bar{P}}}$ is drawn in the red line. The agreement between observations and numerical simulations confirms the theoretical statement.

We have proved that the interplay of a trapping (harmonic oscillator) and an antitrapping (reversed harmonic oscillator) potential generates a novel kind of dispersive shock waves, with the simultaneous presence of annular collapse singularities and a shock-gap enclosed by very intense light barriers.
The use of a thermal medium with a giant Kerr coefficient, the M-Cresol/Nylon solution with $3.5\%$ of nylon concentration, let us access an extremely-nonlinear highly-nonlocal regime and perform accurate experiments with negligible loss. 
We modeled the outcoming dynamics through an advanced theoretical description in rigged Hilbert spaces, by means of time asymmetric quantum mechanics, proving its intrinsic irreversibility.
Our results not only confirm previous studies on the giant nonlinear response of M-Cresol/Nylon, but also disclose fundamental insights on propagation of dispersive shock waves with a singular initial intensity profile.
We believe that this work can be a further step towards a complete description of optical nonlinear phenomena, where inverse scattering transform, Whitham modulation, hydrodynamic approximation and time asymmetric quantum mechanics cooperate in establishing one uniform theory of dispersive shock waves.

We are pleased to acknowledge support from the QuantERA ERA-NET Co-fund 731473 (Project QUOMPLEX), H2020 project grant number 820392, Sapienza Ateneo, PRIN 2015 NEMO, PRIN 2017 PELM, Joint Bilateral Scientic Cooperation CNR-Italy/RFBR-Russia 2018-2020, the NSF Award DMR-1308084, and the National Key R\&D Program of China (2017YFA0303800).

\bibliography{References}
\end{document}